# Vector-based loss functions for turbulent flow field inpainting

Samuel J. Baker[1,*], Shubham Goswami[2], Xiaohang Fang[1,2], and Felix C. P. Leach[1]

*[1]Department of Engineering Science, University of Oxford, United Kingdom*

*[2] Department of Mechanical and Manufacturing Engineering, University of Calgary, Calgary, Canada*

## Abstract

When developing scientific machine learning (ML) approaches, it is often beneficial to embed knowledge of the physical system in question into the training process. One way to achieve this is by leveraging the specific characteristics of the data at hand. In the case of turbulent flows, fluid velocities can be measured and recorded as multi-component vectors at discrete points in space, using techniques such as particle image velocimetry (PIV) or computational fluid mechanics (CFD). However, the vectorised nature of the data is ignored by standard ML approaches, as widely-used loss functions such as the mean-square error treat each component of a velocity vector in isolation. Therefore, the aim of this work is to better preserve the physical characteristics of the data by introducing loss functions that utilise vector similarity metrics. To this end, vector-based loss functions are developed here and implemented alongside a U-Net model for a turbulent flow field inpainting problem, amounting to the prediction of velocity vectors inside large gaps in PIV images. The intention is for the inpainting task to pose a significant challenge for the ML models in order to shed light on their capabilities. The test case uses PIV data from the highly turbulent flow in the well-known Transparent Combustion Chamber III (TCC-III) engine. Loss functions based on the cosine similarity and vector magnitude differences are proposed; the results show that the vector-based loss functions lead to significantly improved predictions of multi-scale flow patterns, while a hybrid (vector and mean-square error) loss function enables a good compromise to be found between preserving multi-scale behaviour and pixel-wise accuracy.

*Keywords:* Turbulent flow, neural network, vector field, loss function, reconstruction, inpainting

## Introduction

Image inpainting is defined as the process of filling in damaged or missing parts of an image, and it has benefitted from the recent explosion in deep learning and computer vision advances [1], [2]. Parallels can be drawn between the images used in computer vision, and the grid- or mesh-based data that often arise in fluid mechanics as a result of experimental methods. For example, a popular flow measurement technique is particle image velocimetry (PIV), in which point-wise velocity measurements are obtained across a flow by tracking the displacement of seeder particles between pairs of images taken over very short time intervals [3]. The result is a set of images that consist of velocity vectors at discrete points (pixels) in the measurement plane, depicting the instantaneous flow patterns. However, the PIV technique is prone to sources of error that can lead to gaps in the resultant vector images, such as irregular seeding density and out-of-plane particle motion, and the challenge is often compounded in the case of internal flow measurement due to complexities such as shadowing (occlusions due to walls or other components), and strong background reflections and light scatter [4], [5]. Incomplete PIV images pose a problem because they limit the level of insight that can be obtained from a flow, have restricted utility in computational fluid dynamics validation pipelines, and are incompatible with modal decomposition and other post-processing techniques [5], [6], [7]. Rectifying PIV errors numerically via inpainting methods has potential to save significant time and money by avoiding experimental re-runs, which is especially valuable for the study of industrially-relevant flows used in prototype development.

Inpainting for fluid mechanics began with various interpolative and modal decomposition-based methods such as Kriging interpolation and the gappy proper orthogonal decomposition [5], [8], [9]. However, the recent development of more powerful techniques that use deep neural networks to capture non-linear relationships has made it feasible to start tackling more challenging tasks such as inpainting with large block gaps [10]. Convolutional neural networks (CNNs) and generative adversarial networks (GANs) have been recently used to successfully make turbulent flow field predictions [11], [12], [13], [14], while physics-informed neural networks (PINNs) [15] have shown promise for creating more generalizable machine learning (ML) models by ensuring that physical properties are respected in the loss function. The present work builds on the latter idea by introducing knowledge of the input data properties into the loss function of a neural network: specifically with the use of vector similarity metrics. A cosine similarity loss has been used in other domains for learning image hashes [16] and cross-modal embeddings [17], [18], but not in such a way as to preserve the characteristics of the input data itself, to the best of the authors' knowledge. Therefore, this paper develops four vector-based loss functions and evaluates their ability to capture turbulent flow patterns compared to a standard mean-square error (MSE) loss.

## Methods

*Dataset*. This work uses a previously developed ML benchmark for turbulent flow machinery named EngineBench [10]. EngineBench consists of PIV data from the Transparent Combustion Chamber (TCC-III) optical engine previously published by the General Motors University of Michigan Automotive Cooperative Research Laboratory [19]. The TCC-III was specifically designed to induce strong turbulence effects via the flat intake and exhaust valves in order to challenge large-eddy simulation (LES) models that were being developed at the time. A variety of physical phenomena can therefore be observed, such as tumble vortex motion, intake jet dynamics, and strong cyclic variations, which can also challenge ML models. Hundreds of gigabytes (GB) of data are available across multiple PIV measurement planes, but a small subset of approximately 0.5 GB is used here, corresponding to 5205 snapshots at five different crank angles (see [10]), in order to simplify the analysis and conform to computer memory constraints. The data subset consists of two-dimensional two-component (2D2C) velocity vectors on the lower-swirl measurement plane which is illustrated in **Error! Reference source not found.**, to ensure the field of view remains constant throughout the full engine cycle as the piston moves. The PIV images therefore depict circular cross-sections of the full 3D cylinder. The field of view is 70 mm in diameter, with final vectors occupying a 50×49 grid (zero-padded to 128×128 for ML model training) with a 1.25 mm vector spacing.

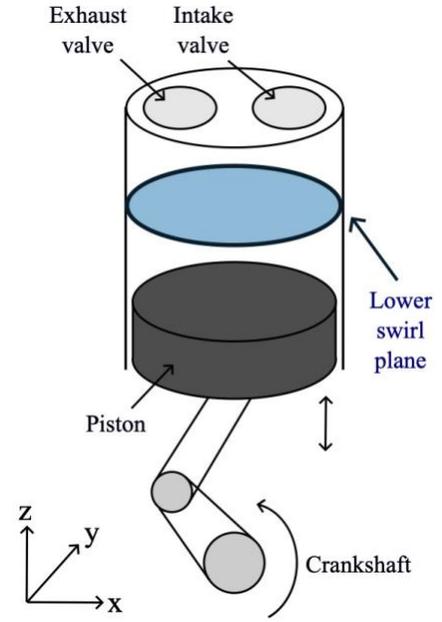

*Figure 1: Schematic of the PIV domain, shown as the lower swirl plane.*

*Model and training process*. The inpainting task is to predict the velocity vectors inside a large gap in the centre of the PIV images, challenging the models by limiting the amount of locally available vector information. For training, data inside a 13×13 square (corresponding to 10% of the visible data) at the centre of the images are set to zero. The gappy snapshots are fed to the ML model, which then makes a prediction that is compared to the original snapshot without the data removed, thus teaching the model to fill the gap. The gap removal process is shown in Figure 3. All of the snapshots at a specific crankshaft phase angle (180 degrees), totalling 1041, are held out for the test set in order to assess generalisability, leaving 4164 PIV images for training. The chosen model is the well-known U-Net [20], due to its high performance in a previous turbulent flow inpainting study [10]. The models were each trained with a

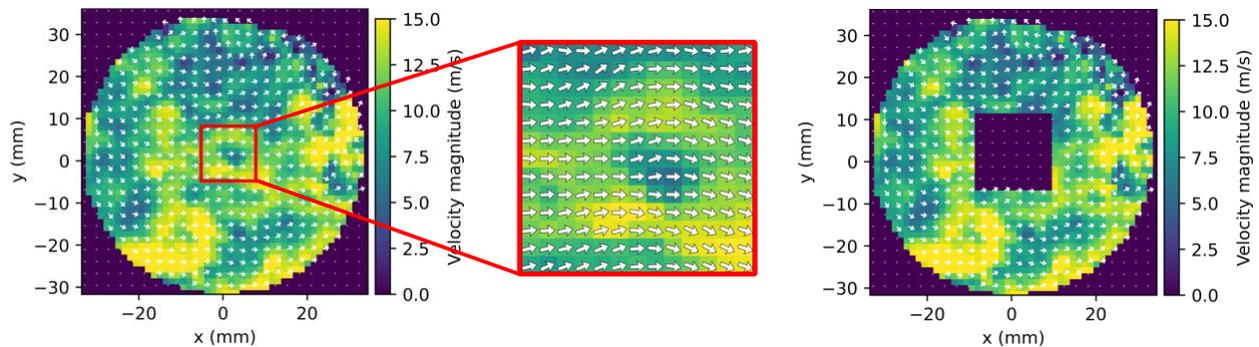

*Figure 2: PIV image from the lower swirl plane of the TCC-III. Vectors are of uniform length, with the colourmap giving the flow speed at each pixel. The central region bounded by the red square is zoomed in for clarity. The vectors in this region are set to zero for to provide the model inputs for training and testing, resulting in a gappy image as shown to the right.*

*Corresponding Author: Samuel J. Baker, samuel.baker@eng.ox.ac.uk*

different loss function for 600 epochs with a learning rate of 0.001, decreasing by a factor of 0.5 every 100 epochs, and an ADAM optimiser with a weight decay of 0.001. All architectural hyperparameters were retained from the project MONAI U-Net [21]. Two metrics are used to evaluate the final predictions on the test set: the normalised L2 error for pixel-wise accuracy, and the Kullback–Leibler (KL) divergence to emphasise larger scale phenomena [10].

*Loss functions.* Four vector-based loss functions are used alongside the MSE for this study, as defined in Table 1. The cosine similarity (also known as the relevance index or RI) and magnitude index (MI) losses are defined such that they scale between 0 for a perfect match, and 1 for perfectly opposite vectors. While the vector loss is constructed of both the cosine similarity and magnitude losses, the hybrid loss is constructed of a combination of the vector and MSE loss functions. The user-defined parameters ($\alpha_1$ and $\underline{\alpha_2}$) were chosen using a grid search of values and chosen based on the best L2 performance on the test data. Therefore, $\alpha_1 = 0.3$ and $\alpha_2 = 0.2$ were used for the entirety of this study.

*Table 1: Vector loss function definitions for input vectors $\boldsymbol{a}$ and $\boldsymbol{b}$, and user-defined constants $\alpha_1$ and $\alpha_2$. The L2 norm is indicated by the $|\ |$ markers.*

| Cosine similarity | Magnitude index (MI) | Vector | Hybrid |
|---|---|---|---|
| $L_{\cos} = \frac{1}{2}\left(1 - \frac{\boldsymbol{a} \cdot \boldsymbol{b}}{|\boldsymbol{a}||\boldsymbol{b}|}\right)$ | $L_{\mathrm{MI}} = \frac{|\boldsymbol{a}-\boldsymbol{b}|}{|\boldsymbol{a}|+|\boldsymbol{b}|}$ | $L_{\mathrm{vec}} = \alpha_1 L_{\cos} + \cdots$ $(1-\alpha_1)L_{\mathrm{MI}}$ | $L_{\mathrm{hyb}} = \alpha_2 L_{\mathrm{vec}} + \cdots$ $(1-\alpha_2)L_{\mathrm{MSE}}$ |

## Results

Quantitative results are provided in **Error! Reference source not found.**, with median values reported alongside a floor (bottom fifth percentile). Figure 3 presents the zoomed-in central sections that correspond to the gap locations, to facilitate closer inspection of the predictions. The cosine loss exhibits the worst performance across the board; indeed, from the figure it is evident that this loss function causes the U-Net to settle on a non-physical solution. The model predictions are very similar across snapshots, consisting of small vectors and near uniform directions. This shows that the cosine loss in isolation is not sensitive enough to the flow dynamics, as the magnitudes of the vectors are not constrained in the cosine similarity equation. Interestingly, the KL divergence flags this behaviour most clearly, reporting 200× higher values than the best model. This demonstrates the importance of a metric that can capture multi-scale phenomena such as the KL divergence, as although the MSE also hints towards the poor performance of the cosine loss, the differences are relatively small considering the level of physical discrepancy between the cosine loss predictions and the ground truth. Meanwhile, the MI loss is influenced by both the magnitudes and the directions of the vectors, and this more complete characterisation of the flow dynamics supports the improved performance with the MI loss across both metrics.

*Table 2: Quantitative results reporting the similarity metrics between the predicted and ground truth flow fields. The floor indicates the bounds for the worst 5% of the results in the test set. The best results for each metric are highlighted in bold.*

| | L2 | | KL | |
|---|---|---|---|---|
| | Median | Floor | Median | Floor |
| Cosine loss | 0.645 | 0.804 | 0.268 | 0.816 |
| MI loss | 0.485 | 0.773 | 0.014 | 0.306 |
| Vector loss | 0.467 | 0.763 | **0.013** | **0.249** |
| Hybrid loss | 0.463 | 0.754 | 0.017 | 0.264 |
| MSE loss | **0.460** | **0.745** | 0.046 | 0.364 |

The vector loss gave similar performances to the MI loss, but with slight improvements caused by an additional emphasis on the vector directions. These losses predicted flow fields with the best KL divergences, at 0.014 and 0.013 for the MI and vector losses respectively. The effect of this can be seen by considering the distributions of velocity magnitudes across the images in Figure 3. In the best-case predictions along the top row of the figure, most of the loss functions exhibit their best performances in flow fields with relatively uniform patterns. However, the best vector loss prediction demonstrates an ability to capture a distribution of vector magnitudes in the form of a high-speed diagonal section with slower regions in the top-right and bottom-left corners. Note that this can come at the cost of point-wise accuracy, with the median vector loss flow field predicting a small vortex in the bottom-left corner that doesn't exist in the true image. Models trained with the MSE loss are less likely to make such an error, with the emphasis on pixel-wise accuracy leading to the strongest L2 error scores in this scenario. However, the median KL divergence is almost four times higher than the vector loss and MI loss models. The effect of this can be seen in the bottom row of Figure 3, where the median MSE loss prediction displays a discrepancy regarding the distribution of velocity magnitudes across the image. The hybrid loss incorporates elements of both the MSE and vector losses and demonstrates a strong and balanced performance. Indeed, the L2 error is just 0.3% higher than the MSE loss, while the KL divergence is 2.7 times lower and significantly closer to the models trained with MI and vector losses.

*Corresponding Author: Samuel J. Baker, samuel.baker@eng.ox.ac.uk

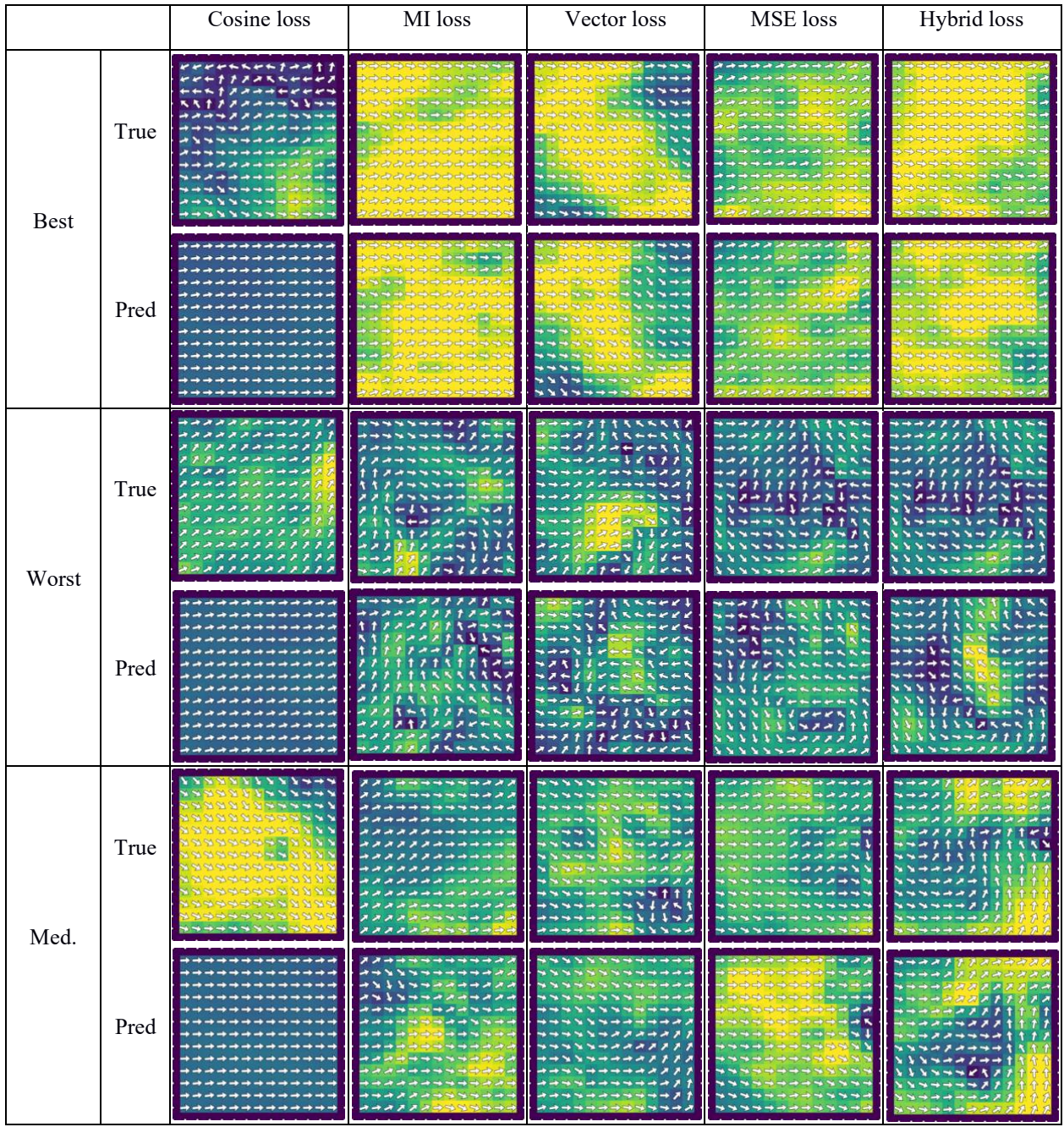

*Figure 3: Corresponding pairs of true and predicted zoomed-in central regions for each loss function. Three cases are reported: the best prediction on the test set according to the L2 error (top row), the worst prediction (middle row), as well as the prediction with the median L2 error (bottom row).*

*Corresponding Author: Samuel J. Baker, samuel.baker@eng.ox.ac.uk

The intention of this test case is to challenge the inpainting models, with the PIV images here characterised by strong turbulence, out-of-plane motion, and wall interactions common to many industrially-relevant internal flows. The best and median performances showcase the potential of the inpainting methods, which are able to predict the general motion of the flow to a fair degree of accuracy even for these large gap sizes. However, the worst-case results can look completely different to the ground truth flows, which limits the utility of a reconstructed flow field set in practice. Inpainting is often described as an ill-posed problem, with many solutions that could potentially fill in the gaps [22]. Therefore, for significant accuracy improvements to be made, it will likely be necessary to further constrain the problem by including additional information such as the pressure or third velocity component. This would provide the models with additional contextual information and possibly facilitate closer adherence to physical conservation laws. Relatedly, another key sensitivity is the size of the gap, with smaller gap sizes also further constraining the solution space. Further work is needed to investigate the interactions between the amount of physical information that is available, and the size of the gaps that a model can be fairly expected to reproduce.

## Conclusions

In this work, four vector-based loss functions were developed for a deep learning inpainting task, with the aim of utilising the vector characteristics of the input PIV data. The investigated loss functions consisted of a cosine similarity loss, a vector magnitude index (MI) loss, a vector loss consisting of a cosine-MI combination, and a hybrid vector-MSE loss. It is found that the MI and vector losses are able to significantly improve the prediction of multi-scale turbulent flow patterns compared to a standard MSE loss as quantified by the KL divergence, at the cost of reduced pixel-wise accuracy. The hybrid loss is able to strike an effective balance between pixel-wise and multi-scale accuracy, which is useful for flows characterised by large multi-scale variations. In this scenario of inpainting 2D2C PIV images of a highly turbulent flow, the best-case and average predictions of the ML models show promising levels of accuracy. However, the worst 5% of the predictions exhibit very large errors compared to the ground truths. This limits the utility of the predicted flow fields; future work should therefore attempt to correlate the performance of the ML models to a physical or textural quantity relevant to the images (such as the combined magnitude relevance index [23]) in order to identify problematic images in advance. Also, it is thought that adding additional constraints to the problem, via inclusion of the third velocity components for example, will aid in the endeavour to improve the 'floor' performances of the models.

## Acknowledgements

Samuel gratefully acknowledges financial support from the EPSRC Doctoral Prize Scheme, grant number EP/W524311/1. Thanks also for the generosity and advice of Miguel A. Mendez at the Von Karman Institute and Michael A. Osborne at Oxford University. Dr. Fang and Dr. Goswami acknowledge support from the NSERC and Alberta Innovation. For the purpose of Open Access, the authors have applied a CC BY public copyright license to any Author Accepted Manuscript (AAM) version arising from this submission.

## References

[1]   O. Elharrouss, N. Almaadeed, S. Al-Maadeed, and Y. Akbari, "Image inpainting: A review," *Neural Process Lett*, vol. 51, pp. 2007–2028, 2020.

[2]   J. Jam, C. Kendrick, K. Walker, V. Drouard, J. G.-S. Hsu, and M. H. Yap, "A comprehensive review of past and present image inpainting methods," *Computer Vision and Image Understanding*, vol. 203, p. 103147, 2021.

[3]   R. J. Adrian and J. Westerweel, *Particle image velocimetry*. Cambridge University Press, 2011.

[4]   C. W. H. Van Doorne and J. Westerweel, "Measurement of laminar, transitional and turbulent pipe flow using stereoscopic-PIV," *Exp Fluids*, vol. 42, pp. 259–279, 2007.

[5]   P. Saini, C. M. Arndt, and A. M. Steinberg, "Development and evaluation of gappy-POD as a data reconstruction technique for noisy PIV measurements in gas turbine combustors," *Exp Fluids*, vol. 57, pp. 1–15, 2016.

[6]   S. J. Baker *et al.*, "Extracting vector magnitudes of dominant structures in a cyclic engine flow with dimensionality reduction," *Physics of Fluids*, vol. 36, no. 2, 2024.

*Corresponding Author: Samuel J. Baker, samuel.baker@eng.ox.ac.uk

[7]     S. Baker *et al.*, "Dynamic Mode Decomposition for the Comparison of Engine In-Cylinder Flow Fields from Particle Image Velocimetry (PIV) and Reynolds-Averaged Navier–Stokes (RANS) Simulations," *Flow Turbul Combust*, no. 0123456789, 2023, doi: 10.1007/s10494-023-00424-3.

[8]     H. Gunes, S. Sirisup, and G. E. Karniadakis, "Gappy data: To Krig or not to Krig?," *J Comput Phys*, vol. 212, no. 1, pp. 358–382, 2006.

[9]     A. Nekkanti and O. T. Schmidt, "Gappy spectral proper orthogonal decomposition," *J Comput Phys*, vol. 478, p. 111950, 2023.

[10]    S. J. Baker, M. A. Hobley, I. Scherl, X. Fang, F. C. Leach, and M. H. Davy, "EngineBench: Flow Reconstruction in the Transparent Combustion Chamber III Optical Engine," *arXiv preprint arXiv:2406.03325*, 2024.

[11]    K. Bao, X. Zhang, W. Peng, and W. Yao, "Deep learning method for super-resolution reconstruction of the spatio-temporal flow field," *Advances in Aerodynamics*, vol. 5, no. 1, p. 19, 2023.

[12]    J. Zhang, J. Liu, and Z. Huang, "Improved deep learning method for accurate flow field reconstruction from sparse data," *Ocean Engineering*, vol. 280, p. 114902, 2023.

[13]    M. Morimoto, K. Fukami, and K. Fukagata, "Experimental velocity data estimation for imperfect particle images using machine learning," *Physics of Fluids*, vol. 33, no. 8, 2021.

[14]    T. Li, M. Buzzicotti, L. Biferale, F. Bonaccorso, S. Chen, and M. Wan, "Multi-scale reconstruction of turbulent rotating flows with proper orthogonal decomposition and generative adversarial networks," *J Fluid Mech*, vol. 971, 2023.

[15]    M. Raissi, P. Perdikaris, and G. E. Karniadakis, "Physics-informed neural networks: A deep learning framework for solving forward and inverse problems involving nonlinear partial differential equations," *J Comput Phys*, vol. 378, pp. 686–707, 2019.

[16]    JT Hoe, KW Ng, T Zhang, CS Chan, YZ Song, and T Xiang, "One loss for all: Deep hashing with a single cosine similarity based learning objective," *Adv Neural Inf Process Syst*, vol. 34, pp. 24286–24298, 2021.

[17]    A. Salvador *et al.*, "Learning cross-modal embeddings for cooking recipes and food images," in *Proceedings of the IEEE conference on computer vision and pattern recognition*, 2017, pp. 3020–3028.

[18]    B. Barz and J. Denzler, "Deep learning on small datasets without pre-training using cosine loss," in *Proceedings of the IEEE/CVF winter conference on applications of computer vision*, 2020, pp. 1371–1380.

[19]    P. Schiffmann, S. Gupta, D. Reuss, V. Sick, X. Yang, and T.-W. Kuo, "TCC-III engine benchmark for large-eddy simulation of IC engine flows," *Oil & Gas Science and Technology–Rev. IFP Energies nouvelle*, vol. 71, no. 1, p. 3, 2016.

[20]    O. Ronneberger, P. Fischer, and T. Brox, "U-net: Convolutional networks for biomedical image segmentation," in *Medical image computing and computer-assisted intervention–MICCAI 2015*, 2015, pp. 234–241.

[21]    M. J. Cardoso *et al.*, "Monai: An open-source framework for deep learning in healthcare," *arXiv preprint arXiv:2211.02701*, 2022.

[22]    D. Pathak, P. Krahenbuhl, J. Donahue, T. Darrell, and A. A. Efros, "Context encoders: Feature learning by inpainting," in *Proceedings of the IEEE Conference on Computer Vision and Pattern Recognition*, 2016, pp. 2536–2544.

[23]    M. Nowruzi, S. Baker, F. Leach, and X. Fang, "Numeric Metrics for Capturing Variations in Flow Fields: An Improvement Towards a Robust Comparison of Vector Fields," *Flow Turbul Combust*, 2025, doi: 10.1007/s10494-025-00637-8.

*Corresponding Author: Samuel J. Baker, samuel.baker@eng.ox.ac.uk